\documentclass[%
 reprint,
 superscriptaddress,
 amsmath,amssymb,
 aps,
prl,
]{revtex4-1}

\usepackage{graphicx}%
\graphicspath{ {Figures/} }
\usepackage{dcolumn}%
\usepackage{bm}%
\usepackage{xcolor}
\usepackage{mhchem}
\newcommand{\avg}[1]{\ensuremath{\left<#1\right>}}
\newcommand{\abs}[1]{\ensuremath{\lvert {#1}\rvert}}

\newcommand{\pds}[2]{\ensuremath{\dfrac{\partial^2 {#1}}{\partial {#2}}}}

\newcommand{\gsl}{\ensuremath{\gamma}}

\newcommand{\stiff}{\ensuremath{\widetilde{\gamma}}}

\newcommand{\ns}{\ensuremath{{n_{\text{s}}}}} %

\newcommand{\vs}{\ensuremath{v_{\text{s}}}}

\newcommand{\ms}{\ensuremath{\mu_{\text{s}}}} %
\newcommand{\ml}{\ensuremath{\mu_{\text{l}}}}
\newcommand{\msl}{\ensuremath{\mu_{\text{sl}}}} %

\newcommand{\tphi}{\ensuremath{\widetilde{\phi}}}
\newcommand{\tphis}{\ensuremath{\widetilde{\phi}_{\text{s}}}}
\newcommand{\tphil}{\ensuremath{\widetilde{\phi}_{\text{l}}}}

\newcommand{\tm}{\ensuremath{T_{\text{m}}}}

\begin{document}

\title{Computing the Tolman length for solid-liquid interfaces}

\author{Bingqing Cheng}
\email{bingqing.cheng@epfl.ch}
 \affiliation{Laboratory of Computational Science and Modeling, Institute of Materials, {\'E}cole Polytechnique F{\'e}d{\'e}rale de Lausanne, 1015 Lausanne, Switzerland}%
 
\author{Michele Ceriotti}
\affiliation{Laboratory of Computational Science and Modeling, Institute of Materials, {\'E}cole Polytechnique F{\'e}d{\'e}rale de Lausanne, 1015 Lausanne, Switzerland}%

\date{\today}

\begin{abstract}
The curvature dependence of interfacial free energy,
which is crucial in quantitatively predicting nucleation kinetics and the stability of bubbles and droplets,
can be described in terms of the Tolman length $\delta$.
For solid-liquid interfaces, however,
$\delta$ has never been computed directly due to various theoretical and practical challenges.
Here we present a general method
that enables the direct evaluation of the Tolman length from atomistic simulations of a solid-liquid planar interface in out-of-equilibrium conditions.
This method works
by first measuring the surface tension from the amplitude of thermal capillary fluctuations of a localized version of Gibbs dividing surface,
and by then computing the free energy difference between the surface of tension and the equimolar dividing surface.
For benchmark purposes,
we computed $\delta$ for a model potential, and compared the results to less rigorous indirect approaches.

\end{abstract}

\maketitle

The Tolman length $\delta$ captures the leading order 
curvature dependence of the interfacial free energy between two phases~\cite{tolman1948consideration,tolman1949effect},
and thus plays a important role in phenomena that involve curved interfaces such as nucleation~\cite{van2009direct} or Ostwald ripening~\cite{voorhees1985theory}.
For many decades, the sign and magnitude of the Tolman length has remained
a source of considerable controversy~\cite{buff1951spherical,van2002determination,lei2005tolman,anisimov2007divergence}, due to the conceptual and practical difficulties involved in measuring or computing it.
For fluid-fluid interfaces,
$\delta$ can be computed from molecular dynamics simulations 
by summing over the local pressure tensor across the phase boundary ~\cite{irving1950statistical,kirkwood1949statistical,van2002determination,van2009direct,zykova2005alkali}.
This same procedure is not directly applicable for solid-liquid interfaces, however,
because the local pressure tensor is ill-defined~\cite{irving1950statistical,blokhuis1992pressure},
and also because the elastic energy stored inside a solid is intertwined with the surface tension.
Due to these theoretical and technical challenges, the value $\delta$ for a solid-liquid interfaces has never been evaluated directly.
Instead it has only ever been used as a fitting parameter that enters the free energy expression for nucleation~\cite{prestipino2012systematic,cheng2017planar}.

The definition of the Tolman length is closely related to the concept of the Gibbs dividing surface, which is an infinitely
thin geometrical surface whose position sensibly coincides with the discontinuity between the two phases. 
The most common way to determine its precise location is to consider a reference system in which the bulk phases extend up to the interface, and has the same value of a chosen extensive quantity (e.g. volume, energy, the sum of order parameters of all atoms) as the actual system, with no excess term associated with the interface
\cite{gibbs1928collected}.
Among all the possible choices for the diving surface,
the surface of tension ($\sigma$-surface), corresponding to the position at which the mechanical definition of tension applies,
is regarded as ``special" because the associated interfacial free energy $\gsl^{\sigma}$ is curvature-independent.
On the other hand, the equimolar dividing surface ($V$-surface) that has no surface excess of volume is commonly used when analyzing nucleation,
because this surface encloses a nucleus that has the same density as the bulk,
which streamlines the formulation of nucleation free energy profile~\cite{cacciuto2003solid,cheng2017planar}.
The Tolman length characterizes the specific free energy 
of a spherical $V$-surface with radius R, such that
$\gsl^{V}(R)=\gsl^{V}(1-2\delta/R+\mathcal{O}(1/R^2))$.
Furthermore, the Tolman length is just the difference between the location of the surface of tension and of the equimolar dividing surface
in the planar limit, i.e. 
~\cite{tolman1948consideration}
\begin{equation}
    \delta = h^V - h^\sigma
    \label{eq:defined}
\end{equation}
where $h^V$ and $h^\sigma$ indicate the height of the two dividing surfaces for the planar interface, and $h=-\infty$ is inside the bulk solid.
Eqn.~\eqref{eq:defined} also implies that
the Tolman length is related to the interfacial free energy difference between the $V$ and the $\sigma$-surfaces by~\cite{tolman1949effect}
\begin{equation}
    \delta= \vs \dfrac{\gsl^{\sigma} - \gsl^{V}}{\msl},
    \label{eq:gettol}
\end{equation}
where $\msl=\ms-\ml$ is the chemical potential difference between the solid and liquid phases,
and $\vs$ is the molar volume of the bulk solid.
$\delta / \vs$ can be interpreted as the adsorption of solid atoms per unit area at the surface
of tension,
and as such Eqn.~\eqref{eq:gettol} directly stems from Gibbs adsorption isotherm~\cite{tolman1948consideration,cheng2015solid}.

Eqn.~\eqref{eq:gettol} provides a recipe for computing the Tolman length using the values of $\gsl^{V}$ and $\gsl^{\sigma}$ at out-of-equilibrium conditions.
A method that employs metadynamics for computing $\gsl^{V}$ away from from the coexistence temperature $\tm$
has only recently become available
~\cite{laio2002escaping,angioletti2010solid,cheng2015solid},
so in the present study we focus on on obtaining $\gsl^{\sigma}$ by applying a capillary wave model.

Even at the macroscopic scale,
a planar solid-liquid interface is not completely flat due to the long wavelength thermal distortions of the interface,
which are generally referred to as capillary waves~\cite{bedeaux1985correlation,hoyt2001method}.
When the extra surface area $\delta s$ is generated due to the capillary waves,
the lattice spacing of the solid is conserved and its elastic energy is unchanged.
Following a mechanical definition of the surface energy,
without heat transfer, change in system size or change in composition of the two phases,
the variation in free energy is entirely captured by the change in surface energy $\delta E = \gsl^{\sigma} \delta s$.
According to the capillary wave model, the surface energy of the fluctuating interface with a height function $h(x,y)$ can be approximated by integrating the angle-dependent surface tension over the interface ~\cite{privman1992fluctuating,hoyt2001method}, i.e.
\begin{equation}
    E_{surf} = \int_s ds \gsl^{\sigma}(\vec{n})
\end{equation}
where $\vec{n}$ is the interface normal vector, and
$ds \approx dx dy (1+(dh/dx)^2/2+(dh/dy)^2/2)$ which is the surface element.
The local curvature of the interface does not enter the equation above as $\gsl^{\sigma}$ is curvature independent by construction.
If each capillary fluctuation mode is in thermal equilibrium,
the interfacial stiffness and the ensemble average of the long wavelength
Fourier components are related via
\begin{equation}
    \avg{\abs{A_h(k_x,k_y)}^2}=
    \dfrac{k_B T}{l_x l_y (k_x^2 \stiff_{11}
    +2 k_x k_y \stiff_{12}
    +k_y^2 \stiff_{22})}
    \label{eq:cfm}
\end{equation}
The $\stiff_{ij} = \gsl^{\sigma} + \pds{\gsl^{\sigma}}{\theta_i \theta_j}|_{\theta_{i,j}=0}$ are the components of stiffness tensor.

Eqn.~\eqref{eq:cfm} is a standard expression for the capillary fluctuation method (CFM)
\cite{hoyt2001method,davidchack2006anisotropic,becker2009atomistic},
which is formally valid at thermal equilibrium and so far has only been employed at the coexistence temperature $T_m$ when the planar interface can be metastable.
~\footnote{CFM has been employed for the water-ice interface formed at undercooled conditions due to premelting~\cite{limmer2014premelting}.}
It was shown in previous work that estimations of the interfacial free energy at $\tm$ for various types of solid-liquid systems using CFM~\cite{hoyt2001method,davidchack2006anisotropic,becker2009atomistic}  are consistent with those obtained using other free energy methods such as metadynamics and thermodynamic integration~\cite{angioletti2010solid,davidchack2000direct,espinosa2016ice,ambler2017solid},
which suggests that the central tenet of CFM -- the mechanical definition of the surface energy -- is valid for solid-liquid interfaces that exhibit capillary fluctuations.
However, for our purposes of using Eqn.~\eqref{eq:gettol},
$\gsl^{\sigma}$ has to be computed at away from $\tm$ so the capillary fluctuation method has to be extended to out-of-equilibrium conditions.
In order to achieve this aim,
as well as to evaluate $\avg{\abs{A_h(k_x,k_y)}^2}$ in Eqn.~\eqref{eq:ak} in a more efficient and accurate manner,
we propose a new and efficient method for locating the fluctuating surface,
which is an extension to the original formulation of Gibbs dividing surface.

Let us consider a solid-liquid system that has $N$ atoms, a box size $\{ l_x, l_y, l_z\}$, and a planar interface perpendicular to the $z$ axis. 
We first introduce an instantaneous order parameter density field %
\begin{equation}
    \tphi(x,y,z) = \sum_{i=1}^{N}
    \phi_i g(x-x_i)g(y-y_i)g(z-z_i),
    \label{eq:phasefield}
\end{equation}
where $(x_i,y_i,z_i)$ denote the coordinate of the $i$-th atom, $\phi_i$ is an atom-centered order parameter that is able to discriminate between the atoms that belong to each of the two different phases,
and $g$ is a normalized kernel function which is chosen to be a Dirac delta function in this case.
The zero-excess condition that defines the height of the interface $h(x,y)$ at any point can then be written in terms of a line integral of the phase field (Eqn.~\eqref{eq:phasefield}) along the $z$ axis over a range that contains the interface
\begin{equation}
     \int_{0}^{l_z} dz \tphi(x,y,z) = 
     \int_0^{h(x,y)} dz \tphis
     + \int_{h(x,y)}^{l_z} dz \tphil,
     \label{eq:gibbs}
\end{equation}
Here $\tphis$ and $\tphil$ indicate the density field inside the bulk solid and liquid phases, respectively.
Eqn.~\ref{eq:gibbs} can be seen as an extension of the planar Gibbs dividing surface for the whole system~\cite{cheng2015solid,cheng2016bridging} that is restricted to an infinitesimally thin domain centered around $(x,y)$.
Combining Eqn.~\eqref{eq:phasefield} and~\eqref{eq:gibbs}, one can write
\begin{equation}
    \int_0^{h(x,y)} dz \tphis
     + \int_{h(x,y)}^{l_z} dz \tphil=
    \sum_{i=1}^N \phi_i  g(x-x_i)g(y-y_i).
    \label{eq:h}
\end{equation}

As discussed in the SI~\cite{SI}, by taking a 2-D Fourier expansion for both sides of
Eqn.~\eqref{eq:h} over the whole cross section of the simulation box $\{l_x,l_y\}$
and performing an ensemble average for the amplitude of each Fourier mode, 
one obtains
\begin{multline}
        (\avg{\tphis}-\avg{\tphil})^2 \avg{\abs{A_h(k_x,k_y)}^2}\\
        + \avg{\abs{A_s(\avg{h};k_x,k_y)}^2}
        + \avg{\abs{A_l(l_z-\avg{h};k_x,k_y)}^2}\\
        = \avg{\left[\dfrac{1}{ l_x l_y}  
   \sum_{i=1}^N  \phi_i \exp{(-\mathrm{i}k_x x_i - \mathrm{i} k_y y_i)}\right]^2},
   \label{eq:ak}
\end{multline}
by assuming that the bulk fluctuations and the surface fluctuations are mutually independent.
In Eqn.~\eqref{eq:ak}, 
$\avg{\tphis}$ and $\avg{\tphil}$ are the averaged values from the bulk fields.
$A_s(\avg{h};k_x,k_y)$ and $A_l(l_z-\avg{h};k_x,k_y)$ are the Fourier coefficients characterizing the bulk fluctuations, respectively for a slab of bulk solid that has a cross section $\{ l_x, l_y\}$
and thickness $\avg{h}$, and 
for a bulk liquid with the dimensions $\{ l_x, l_y, l_z-\avg{h}\}$.
The average amplitudes of these bulk quantities can all be evaluated separately from simulations of the bulk phases using a simulation box of the same cross section as the solid-liquid system,
using expressions such as
\begin{multline}
\avg{\abs{A_s(\avg{h};k_x,k_y)}^2} =\\
\avg{\left[\dfrac{1}{ l_x l_y}  
   \sum_{i=1}^{N_s}  \phi_i \exp{(-\mathrm{i}k_x x_i - \mathrm{i} k_y y_i)}H(\avg{h}-z_i)H(z_i)\right]^2},
\end{multline}
where $H(\ldots)$ is the Heaviside function and $N_s$ is the number of atoms in the bulk solid system.
As the right hand side of Eqn.~\eqref{eq:ak}
can also be evaluated directly from the snapshots of atomic coordinates for the solid-liquid system in molecular dynamics simulations,
the only remaining term $\avg{\abs{A_h(k_x,k_y)}^2}$ that enters the CFM expression (Eqn.~\eqref{eq:cfm}) can be determined.
Note that when there are two parallel planar interfaces in the system,
one can simply modify Eqn.~\eqref{eq:ak} by adding a factor of two to the term $\avg{\abs{A_h(k_x,k_y)}^2}$.

The zeroth Fourier mode of the height function $A_h(0,0)$,
corresponds to the average height of the fluctuating interface.
$
\avg{h} = \int_0^{l_x} dx \int_0^{l_y} dy h(x,y)/ l_x l_y
$
corresponds to the position of the conventional planar Gibbs dividing surface of the whole system that has zero surface excess for the extensive quantity $\Phi=\sum_{i=1}^{N}\phi_i$, since
\begin{multline}
\Phi=
\int_0^{l_x} dx \int_0^{l_y} dy \left[\int_0^{\avg{h}} \tphis dz
  + \int_{\avg{h}}^{l_z} \tphil  dz\right].
  \label{eq:Phi}
\end{multline}
As extensively discussed in Ref.~\cite{tolman1948consideration,cheng2015solid},
the location of this planar dividing surface $\avg{h}$ determines
its surface absorption and thus affects its interfacial free energy.
Meanwhile, the magnitudes of all other Fourier modes with non-zero frequencies ($A_h(k_x,k_y)$) do not change the proportion between the solid and the liquid atoms in the system, and thus do not affect the surface absorption.

Realizing that $\Phi$ only depends on the zeroth capillary fluctuation mode $\avg{h}$ (Eqn.\eqref{eq:Phi})
makes it possible to extend CFM to conditions that are away from the coexistence temperature $\tm$.
Under such conditions, the driving force for interface migration stems from the chemical potential difference between the metastable and the stable phases,
and an umbrella potential can be introduced to the system to counter-balance this force and to pin the interface \cite{pedersen2013computing}.
Taking the actual Hamiltonian of the system to be $\mathcal{H}(\textbf{q})$,
the biased Hamiltonian can be expressed as
\begin{equation}
    \mathcal{H}_{biased}(\textbf{q}) = \mathcal{H}(\textbf{q})
    + \dfrac{\alpha}{2} \left( \Phi - \bar{\Phi}\right)^2.
\end{equation}
As the bias potential does not act on any capillary modes other than the zeroth mode $\avg{h}$,
the equipartition theorem still holds for other non-zero frequencies and Eqn.~\eqref{eq:cfm} is thus still valid.

Another subtle point is that, 
at a reasonably large length scale, the height function $h(x,y)$ of the fluctuating interfaces defined by different order parameters are parallel to each other,
and the magnitudes for long wavelength Fourier modes are identical. 
As extensively discussed in the SI~\cite{SI},
for each $(k_x,k_y)$ wave vector that is smaller than a certain cutoff,
the Fourier amplitude and the corresponding value of $\stiff$ 
are independent from the choice of the order parameter that is used to define the fluctuating dividing surface.
Previous studies that employ different order parameters and even distinct criteria for locating the interface also arrived at consistent estimations for $\stiff$ and $\gsl^{\sigma}$ at $\tm$~\cite{hoyt2001method,davidchack2006anisotropic,becker2009atomistic,baldi2017extracting}.

We simulated the solid-liquid planar interfaces for 
a simple but realistic Lennard-Jones system~\cite{davidchack2003direct,angioletti2010solid,benjamin2014crystal}.
The NPT ensemble was employed throughout with the stochastic velocity rescaling thermostat~\cite{bussi2007canonical}.
A Nose-Hoover barostat was used along the $z$ axis which was set up to be perpendicular to the interface,
The dimensions of the supercell along $x$ and $y$ are commensurate with the  equilibrium lattice parameters.
We simulated planar interfaces along the $\avg{100}$ and $\avg{110}$ crystallographic directions of the \emph{fcc} lattice,
and the system size were 256000 and 315392 atoms, respectively.
Umbrella sampling~\cite{torrie1977nonphysical} simulations were run at temperatures of
$0.56$, $0.58$, $0.60$ and $\tm=0.6178$,
using the collective variable $\Phi=\sum_i S(\kappa(i))$ that was described in Ref.~\citenum{cheng2015solid}.
Fast implementation of this simulation setup was made possible by the flexibility of the PLUMED code~\cite{tribello2014plumed} in combination with LAMMPS~\cite{plim95jcp}.
The supplemental information contains annotated sample input files~\cite{SI}.

By combining Eqn.~\eqref{eq:cfm} and Eqn.~\eqref{eq:ak},
the interfacial stiffness $\stiff$ can be extracted from each thermal capillary fluctuation mode during post processing,
and a Python notebook that evaluates $\avg{\abs{A_h(k_x,k_y)}^2}$ and $\stiff$ directly from snapshots of the configurations of a solid-liquid system is include in the SI~\cite{SI}.
We have also included a convergence test of extracting $\stiff$ using different order parameters,
as well a detailed discussion of how to select the order parameter as well as a cutoff for the wave vector
to ensure an efficient and accurate evaluation of $\stiff$~\cite{SI}.

\begin{figure}
\includegraphics[width=0.5\textwidth]{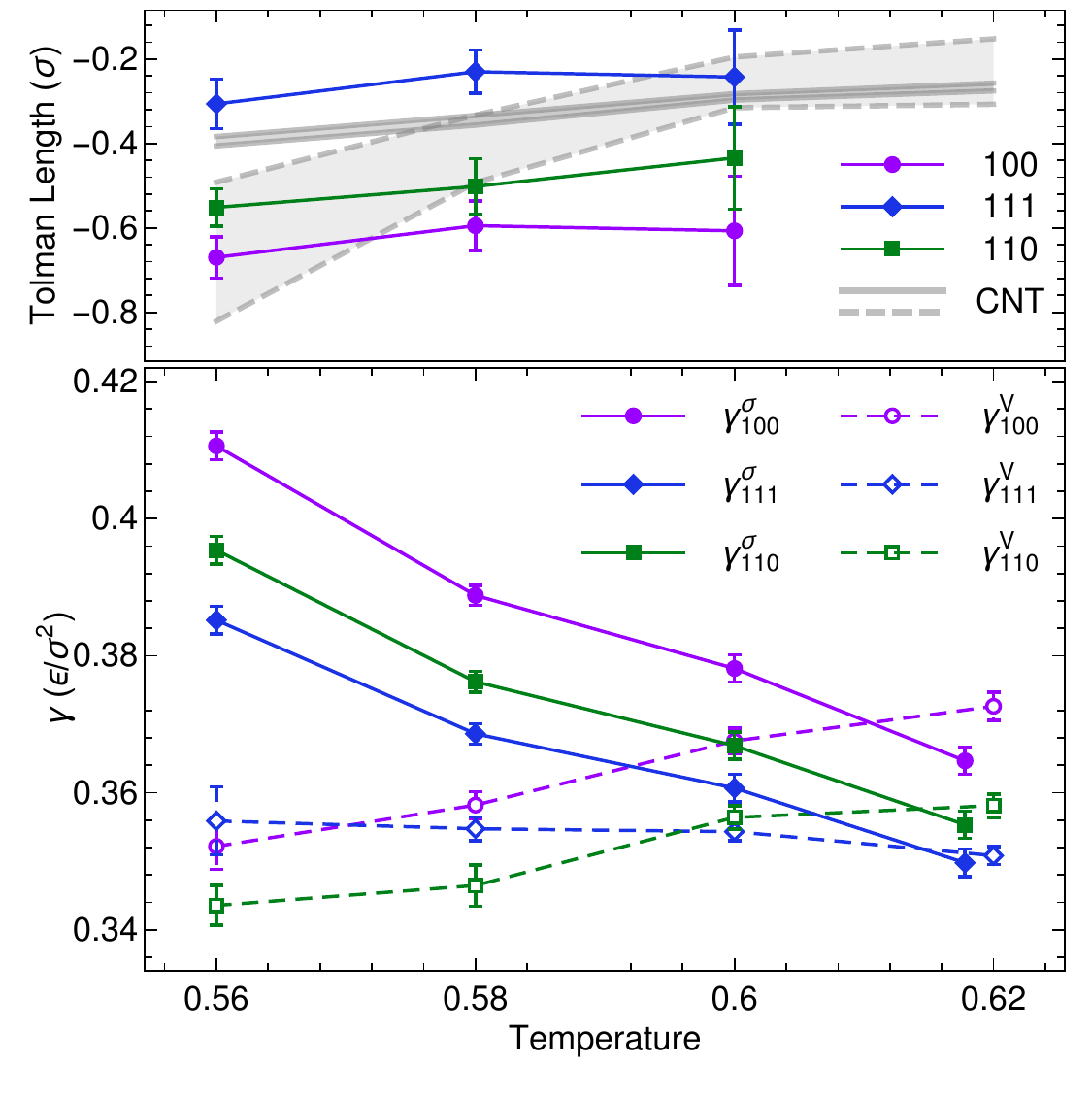}
\caption{
The bottom panel shows the the planar interfacial free energies of the surface of tension ($\gsl^{\sigma}$) and the equimolar dividing surface ($\gsl^{V}$) for three different lattice directions at different temperatures ($\tm=0.6178$).  
The data for $\gsl^V$ were obtained from Ref.~\cite{cheng2015solid}.
The purple, blue and green symbols in the top panel indicate the predictions of the Tolman length for the $\avg{100}$, $\avg{111}$ and $\avg{110}$ solid-liquid interfaces.
The dashed gray band shows indirect estimates of the orientation-averaged Tolman length of a solid nucleus from homogeneous nucleation simulations,
where the Tolman length,  the chemical potential difference and the interfacial free energy were treated as three fitting parameters in a classical nucleation theory expression~\cite{cheng2017planar}.
The solid gray band indicates another set of results from fitting nucleation free energy profiles,
where $\delta$ is used as the sole parameter, and the interfacial free energies and chemical potentials are obtained elsewhere from independent planar interface simulations~\cite{pedersen2013computing,cheng2015solid}.
Statistical errors of the mean estimations are indicated using either error bars or band widths.}
\label{fig:se}
\end{figure}

For the $\avg{100}$ or $\avg{110}$ interfaces, we first extracted the interfacial stiffness tensors $\stiff_{11}$ and $\stiff_{22}$ using Eqn.~\eqref{eq:ak} and Eqn.~\eqref{eq:cfm} at each temperature from 10 independent biased molecular dynamics runs.
After that, we expanded the stiffness tensors in cubic harmonics, which are consistent with the symmetry of the \emph{fcc} crystal, so as to extract
the orientation-dependent interfacial free energy $\gsl^{\sigma}(\vec{n})$~\cite{becker2009atomistic}.
In Figure~\ref{fig:se} we plot the interfacial free energy of the surface of tension for the $\avg{100}$, $\avg{111}$ and $\avg{110}$ interfaces.
We also plot the free energies of planar equimolar dividing surfaces $\gsl^V$ that were obtained from our previous metadynamics simulations~\cite{cheng2015solid}.
Unsurprisingly, the value, temperature dependence, and anisotropy among the principal orientations of $\gsl$ vary considerably with the choice of the Gibbs dividing surface.  
\footnote{In theory, at $T_m$, $\gsl^{\sigma}=\gsl^{V}$, but in reality we notice a small difference between them. 
This is due to the fact that $\gsl^{V}$ was computed using metadynamics simulations with small supercells comprising about 1200 atoms, 
which implies a small finite-size effect that increases interfacial free energies~\cite{angioletti2010solid}, and also shifts $\tm$ to  $0.62$~\cite{cheng2015solid}.
In order to compensate for this small finite-size effect, we shifted vertically the values of $\gsl^{V}$ so that $\gsl^{\sigma}(T_m)=\gsl^{V}(T_m)$ before computing the Tolman length using Eqn.~\eqref{eq:gettol}.}

The purple, blue and green symbols in the top panel of Figure~\ref{fig:se} indicate the estimates of the Tolman length for the $\avg{100}$, $\avg{111}$ and $\avg{110}$ interfaces.
There is significant anisotropy in $\delta$ among the three crystal orientations, and little temperature dependence.
The negative value of $\delta$ suggest that the equimolar dividing surface is closer to the bulk solid than the surface of tension, and that the interfacial energy of a curved equimolar surface $\gsl^V(R)$ has a positive curvature dependence.
For comparison, in Figure~\ref{fig:se} we also plot the estimate of the orientation-averaged Tolman length for three dimensional solid nucleus from previous homogeneous nucleation simulations~\cite{cheng2017planar}.
These values were obtained by fitting
nucleation free energy profiles with a classical nucleation theory expression with a Tolman correction term:
\begin{equation}
 G(\ns(V))= \msl \ns(V) + \gsl^{V} \Omega \vs^{\frac{2}{3}} \ns^{\frac{2}{3}}(V)
 (1- \epsilon \ns^{-\frac{1}{3}}(V)),
 \label{eq:cntv2}
\end{equation}
where $\epsilon = (32\pi/3)^{\frac{1}{3}}\vs^{-\frac{1}{3}}\delta$, and $\Omega$ is a geometrical constant~\cite{cheng2017planar}.
One can treat $\msl$, $\gsl^{V}$ and $\delta$ as three fitting parameters,
and the dashed gray band in Figure~\ref{fig:se} shows the indirect estimation of the Tolman length $\delta$ using this approach.
One can also obtain $\msl$ and $\gsl^{V}$ independently from other metadynamics simulations of planar interfaces~\cite{pedersen2013computing,cheng2015solid} and just use $\delta$ as the only fitting parameter.
$\delta$ estimated in such a way is plotted with the solid gray band in Figure~\ref{fig:se}.
Overall, the estimations of the Tolman length from all these three different methods agree.
This agreement not only corroborates the present framework and method,
but also indicates that the CNT expression with the three parameters (the chemical potential difference $\msl$, the interfacial free energy $\gsl^{V}$ and the Tolman length $\delta$) can accurately describe the free energy profile for homogeneous nucleation of the Lennard-Jones crystal from its melt.

In summary, we presented a thermodynamic framework that enables a direct evaluation of the Tolman length for planar solid-liquid interfaces in atomistic simulations.
This framework is the coronation of an effort to streamline the study of nucleation by means of atomistic simulations, and relies on (i) the rigorous definition of the Gibbs dividing surface based on an atomic-scale order parameter and its fluctuations~\cite{cheng2015solid,cheng2016bridging}, which we have used here to formulate an elegant and efficient version of the capillary fluctuation method that does not require the explicit geometric location of the dividing surface; (ii) the use of metadynamics to compute the planar-interface surface energy for the equimolar dividing surface  $\gsl^{V}$ in out-of-equilibrium conditions~\cite{angioletti2010solid,cheng2015solid}; (iii) the calculation of the mechanical surface tension $\gsl^{\sigma}$  based on a capillary fluctuation analysis away from coexistence conditions. 
Based on these theoretical advances, we computed $\gsl^{\sigma}$ and the Tolman lengths $\delta$ of the solid-liquid planar interfaces of three principal crystal directions for a model system.
In this case, we find that the values of $\delta$ that we obtained by evaluating it directly are in good agreement with the ones obtained by fitting homogeneous nucleation free energy profiles using CNT expressions.
The framework we presented opens the door to a rigorous determination of the Tolman length $\delta$ and the free energy of the surface of tension $\gsl^{\sigma}$ for various physical systems, and for different classes of homogeneous and heterogeneous phase transitions. 
The ability to compute these quantities is crucial both in verifying the consistency of classical nucleation theory for a given problem, and obtaining a quantitative prediction of nucleation rates by means of atomistic modelling.

\begin{acknowledgements}
We thank Alessandro Laio and Erio Tosatti for insightful comments and stimulating discussions.
We also thank Gareth Tribello, Eduardo Sanz,
and Rocio Semino
for critically reading the manuscript.
MC and BC would like to acknowledge funding from the Swiss National Science 
Foundation (Project ID 200021-159896) and generous allocation of CPU time by
CSCS under Project ID s787.
\end{acknowledgements}

\end{document}